\documentclass[10pt]{article}
\usepackage{euscript,amssymb}

\font\msbm=msbm10
\def\RR{\hbox{\msbm R}}

\def\ZZ{\hbox{\msbm Z}}

\catcode`\@=11
\def\lesssim{\mathrel{\mathpalette\vereq<}}
\def\vereq#1#2{\lower3pt\vbox{\baselineskip1.5pt \lineskip1.5pt
\ialign{$\m@th#1\hfill##\hfil$\crcr#2\crcr\sim\crcr}}}

\def\Let@{\relax\iffalse{\fi\let\\=\cr\iffalse}\fi}
\def\vspace@{\def\vspace##1{\crcr\noalign{\vskip##1\relax}}}
\def\multilimits@{\bgroup\vspace@\Let@
 \baselineskip\fontdimen10 \scriptfont\tw@
 \advance\baselineskip\fontdimen12 \scriptfont\tw@
 \lineskip\thr@@\fontdimen8 \scriptfont\thr@@
 \lineskiplimit\lineskip
 \vbox\bgroup\ialign\bgroup\hfil$\m@th\scriptstyle{##}$\hfil\crcr}
\def\Sb{_\multilimits@}
\def\endSb{\crcr\egroup\egroup\egroup}
\def\Sp{^\multilimits@}

\newcommand{\be}[1]{\begin{equation}\label{#1}}
\newcommand{\ee}{\end{equation}}
\newcommand{\ba}[1]{\begin{eqnarray}\label{#1}}
\newcommand{\ea}{\end{eqnarray}}
\newcommand{\rf}[1]{(\ref{#1})}
\newcommand{\nn}{\nonumber}
\newcommand{\const}{\mbox{\rm const}}

\begin{document}

\author{Mariam Bouhmadi--L\' opez\dag\footnote{e-mail:
mbouhmadi@imaff.cfmac.csic.es} \ and Alexander
Zhuk\dag\footnote{e-mail: ai$\_$zhuk@imaff.cfmac.csic.es
\newline
on leave from: Department of Physics, University of Odessa, 2
Dvoryanskaya St., Odessa 65100, Ukraine}\\ \dag
 Instituto de Matem\'aticas y F\'{\i}sica
Fundamental, CSIC, \\ C/ Serrano 121, 28006 Madrid, Spain }

\title{Comments on conformal stability of brane-world models}

\date{July 26, 2001}
%\date{}
\maketitle

\abstract{ The stability of 5--D brane-world models under
conformal perturbations is investigated. The analysis is carried
out in the general case and then it is applied to particular
solutions. It is shown that models with the Poincar\' e and the de
Sitter branes are unstable because they have negative mass squared
of gravexcitons whereas models with the Anti de Sitter branes have
positive gravexciton mass squared and are stable. It is also shown
that 4--D effective cosmological and gravitational constants on
branes as well as gravexciton masses undergo hierarchy: they have
different values on different branes.}

\bigskip

\hspace*{0.950cm} PACS number(s): 04.50.+h, 98.80.Hw

%%%%%%%%%%%%%%%%%%%%%%%%%%%%%%%%%%%%%%%%%%%%%%%%%%%%%%%%%%%%

%%%%%%%%%%%%%%%%%%%%%%%%%%%%%%%%%%%%%%%%%%%%%%%%%%%%%%%%%%%%

\section{Introduction}
\setcounter{equation}{0}

\bigskip

It is well known that part of any realistic multidimensional model
should be a mechanism for extra dimension stabilization. This
problem was a subject of numerous investigations. In the standard
Kaluza-Klein approach cosmological models are taken in the form of
warped product of Einstein spaces as internal spaces.
Corresponding warp (scale) factors are assumed to be functions of
external (our) space-time. If these scale factors are dynamical
functions then it results in a variation of the fundamental
physical constants. To be in agreement with observations, internal
spaces should be compact, static (or nearly static) and less or
order of electro-weak scale (the Fermi length). The stability
problem of these models with respect to conformal perturbations of
the internal spaces was considered in detail in our paper
\cite{GZ}. It was shown that stability can be achieved with the
help of an effective potential of a dimensionally reduced
effective 4--D theory. Small conformal excitations of the internal
spaces near minima of the effective potential have the form of
massive minimal scalar fields developing in the external
space-time. These particles were called gravitational excitons
(gravexcitons). Their physical meaning can be easily explained
with the help of a simple 3--D model where 2--D spatial part has
the cylindrical topology: $S^1\times R^1$. Here, $S^1$ plays the
role of the compact internal space and $R^1$ describes 1--D
external space. Let us suppose that the size of $S^1$ is
stabilized near some value by an effective potential. Then,
conformal excitation of $S^1$ near its equilibrium position
results in waves running along the cylinder (along $R^1$). Thus,
any 1--D observer living on the cylinder (on $R^1$) will detect
these oscillations as massive scalar fields. Obviously, this
effect takes place for any multidimensional cosmological model
with compact internal spaces. In general, it does not depend on
presence or absence of branes in models. Masses of gravexcitons
and equilibrium positions for the internal spaces depends on the
form of the effective potential (on concrete topology and matter
content of the model).

Recently \cite{ADD,AADD,ADM}, it was realized that it is not
necessary for extra dimensions to be very small. They can be
enlarged up to sub-millimeter scales in such a way that the
Standard Model fields are localized on a 3--brane with thickness
of the electro-weak (or less) length in the extra dimensions
whereas gravitational field can propagate in all multidimensional
(bulk) space. It gives possibility for lowering of the
multidimensional fundamental gravitational constant down to the
TeV scales (therefore this approach often call as TeV gravity
approach). Cosmological models in this approach are topologically
equivalent to the standard Kaluza-Klein one. Problem of their
stability against conformal perturbations of additional dimensions
was considered in papers \cite{ADM,ADKM}. A comparison of old and
this new approaches from the point of view of conformal stability
was given in paper \cite{GZII}. In papers \cite{ADM,ADKM}
conformal excitations of the additional dimensions near minimum
position of the effective potential were called radions (to our
knowledge, the first time that the term radion appeared it was in
\cite{ADM}). However, we prefer to call such particles
gravexcitons because, first, from the point of priority and,
second and it is the most important, the term radion is widely
used now in the brane-world models in different content.

The brane-world models are motivated by the strongly coupled
regime of $E_8\times E_8$ heterotic string theory which is
interpreted as M-theory on an orbifold $\RR^{10}\times S^1/\ZZ_2$
with a set of $E_8$ gauge fields at each ten-dimensional orbifold
fixed plane. After compactification on a Calabi-Yau three-fold and
dimensional reduction one arrives at effective $5-$dimensional
solutions which describe a pair of parallel 3-branes with opposite
tension, and located at the orbifold planes \cite{HW}. For these
models the $5-$dimensional metric contains a $4-$dimensional
metric component multiplied by a warp factor which is a function
of the additional dimension. A cosmological solution of this type
with flat 4--D branes (which we shall refer as Poincar\' e branes)
was obtained in paper \cite{RS}. This model was generalized in
numerous publications to the cases of bent branes in models with 5
and more dimensions and with single or many branes. In paper
\cite{CGRT} it was stressed the necessity of stabilization the
distance between branes to get conventional cosmology on branes.
Here, the radius (scale factor) of the extra dimension was called
radion. But this definition is not very precise. We should note
that there is a confusion in the literature concerning the term
radion: quite different forms of the metric perturbations of the
brane-world models were called radions. However, in paper
\cite{CGR} it was clearly shown and strictly emphasized that
radions describe relative distance between branes (see also papers
\cite{BDL,CF}). It demonstrates the main difference between
gravexcitons and radions: gravexcitons describe conformal
excitations of geometry (more particular - conformal excitations
of the additional dimensions) whereas radions describe relative
motion of branes. Obviously, gravexcitons can exist in model where
branes are absent at all or in models with a single brane and vice
versa radions can exist in absence of gravexcitons. The latter
situation can be realized for example in the TeV scale approach
where branes can move relatively with respect each other due to
interaction between them "sliding" on background fixed geometry
(gravexcitons are absent). Branes are considered here as "probe
bodies" moving in the background geometry. Nevertheless, in the
brane-world models gravexcitons and radions are closely connected
with each other (and this is the main reason for the confusion
between them). Here, branes are 4--D surfaces along which
different 5--D bulk solutions are gluing with each other. In this
case positions of branes fix the shape and size of the
geometry\footnote{In our paper we shall consider the case of
compact with respect to additional dimension brane-world
models.\label{1}} and relative motion of branes results in
conformal changes of the geometry. Thus, it is natural for such
models to expect that solutions stable against radions are stable
against gravexcitons and vice versa. Obviously, direct comparison
of stability against gravexcitons and radions has only sense in
models with two (and more) branes where radions exist.

The radion stabilization problem was investigated in a number of
papers devoted to the brane-world models. It was shown in
particular, that in the case of solutions with two de Sitter
branes (bent branes with 4--D effective positive cosmological
constants) and two Anti de Sitter branes (bent branes with 4--D
effective negative cosmological constants) radions have negative
and positive mass squared respectively. Thus, the former solution
is unstable but the latter one is stable against radions. In our
paper we find, first, that the model with one de Sitter brane is
unstable against gravexcitons and, second, that the model with one
or a number of parallel Anti de Sitter branes (connected with each
other via wormhole throats) is stable under conformal excitations.
In the case of Randall-Sundrum solution \cite{RS} with two
Poincar\' e branes, radions have zero mass \cite{BDL,CF}. From the
Particle Physics point of view, such particles do not lead to
instability. However, as it is well known, such ultra-light scalar
fields, originated from the extra dimensions, produce a number of
cosmological problems connected with flatness of their effective
potential. For example, in the homogeneous case any small
excitations near an equilibrium position (which can be chosen
arbitrary for the flat potential) have a runaway behaviour if we
do not take into account the friction term due to the cosmological
expansion. However, such dynamical stabilization is very delicate
problem (see e.g. \cite{GZ(CQG)}) and needs a separate
investigation for each case. Exactly this kind instability for
radions in the model with two Poincar\' e branes was found in
\cite{BDL}. It was shown here that a small departure from the
equilibrium position results in either a colliding of the two
branes or a runaway behaviour. In our paper we show that models
with two as well as one Poincar\' e brane are also unstable under
conformal perturbations. It is necessary to stress that in papers
\cite{BDL,CF} the analysis was performed in the Brans--Dicke
frame. In our paper the problem of stability is investigated in
the Einstein frame. Obviously, if models are stable in one frame
they are stable in another one because both of frames coincide in
the equilibrium position. However, exact form of the dynamical
behaviour (time dependence) near the equilibrium position can
depend on the frame. In paper \cite{Od} it is explicitly discussed
the role of conformal transformations and it is shown that some
solutions of the brane-world models exist in one frame but are
absent in the other one. Therefore, the equivalence between these
two frames depends on the concrete discussed problem and in some
cases is a matter of a delicate investigation.

%It was shown in particular, that Randall-Sundrum solution
%\cite{RS} with two Poincar\' e branes as well as solution with a
%pair of the de Sitter branes (bent branes with 4--D effective
%positive cosmological constants) are unstable whereas solution
%with a pair of the Anti de Sitter branes (bent branes with 4--D
%effective negative cosmological constants) is stable
%\cite{BDL,CF}. This conclusion coincides with the results of our
%paper.

In papers \cite{BDL,CF}, mentioned above, the authors conclusions
concerning the radion stability/instability were obtained for the
brane-world models where matter in bulk as well as on branes is
absent (correctly speaking, it is considered there in its simplest
form as bulk cosmological constant and "vacuum energies" on
branes). In this case, only the pair of the Anti de Sitter branes
are really stable. However, it was observed that inclusion of
matter can stabilize radions for different types of branes. It can
be done with the help of bulk scalar field \cite{CGRT},
\cite{GW}--\cite{KKS}, perfect fluid on branes \cite{GNS} and the
Casimir effect between branes \cite{GPT,HKP}.

Some specific forms of instability in the brane-world solutions
were observed in papers \cite{BDB,ALS}. It was shown that single
Poincar\' e brane is unstable under small perturbations of the
brane tension\footnote{Here it was mentioned about instability of
the single brane Randall-Sundrum solution under homogeneous
gravitational perturbations. In our paper we show that this model
is unstable also under conformal excitations.\label{Durer}}
\cite{BDB} and single de Sitter brane is unstable against thermal
radiation \cite{ALS}.

The main goal of our present comments consists in investigation of
5--D brane-world stability against conformal perturbations. First,
we elaborate a method to study the stability for a large class of
solutions and obtain general expressions for 4--D effective
cosmological constants on branes and masses of gravitational
excitons. Then we apply this method to a number of well known
solutions. In particular, we find that models with the Poincar\' e
and the de Sitter branes are unstable because they have negative
mass squared of gravexcitons whereas models with the Anti de
Sitter branes have positive gravexciton mass squared and are
stable under conformal perturbations. We show also that 4--D
effective cosmological and gravitational constants on branes as
well as gravexciton masses undergo hierarchy: they have different
values on different branes (if different branes have different
warp factors).

The paper is organized as follows. In Section II %\ref{setup}
we explain the general setup of our model, perform dimensional
reduction of the brane-world models to an effective 4--D theory in
general case
%describing a large class of models
and apply this procedure to a number of well known solutions. In
Section III
%\ref{stability}
we elaborate a method of the investigation of brane-world solution
stability against conformal perturbations and apply it to
particular solutions considered in Section II.
%\ref{setup}.
%and show, in particular, that solutions with the
%Poincar\' e and de Sitter branes are unstable whereas the Anti de
%Sitter brane solutions are stable.
%for stable solutions
Here we show also that physical masses of gravexcitons undergo
hierarchy on different branes. The brief Conclusions of the paper
are followed by three appendices. In Appendix A we present useful
expressions for the Ricci tensor components and scalar curvature
in the case of block-diagonal metrics. Some useful formulas of the
conformal transformation are summarized in Appendix B. In Appendix
C we show that the results of the paper do not change if only
additional dimension undergoes conformal perturbations: we arrive
here to the same 4--D effective theory and the same gravexciton
masses as in the case of total geometry conformal perturbations.
This provides an interesting analogy between gravity and an
elastic media where the eigen frequencies of an elastic body
oscillations do not depend on the manner of excitation.

%%%%%%%%%%%%%%%%%%%%%%%%%%%%%%%%%%%%%%%%%%%%%%%%%%%%%%%%%%%%%%
\section{Model and general setup: dimensional reduction of
brane-world models\label{setup}} \setcounter{equation}{0}

\bigskip

We consider 5--D cosmological models on a manifold $M^{(5)}$ which
is divided on $n$ pieces by $n-1$ branes: $M^{(5)} = \cup_{i=1}^n
M^{(5)}_i$. Branes are 4--D hypersurfaces $ r=r_i= \const \, ,
\quad i= 1,\ldots ,n-1$, where $r$ is an extra dimension. Each
brane is characterized by its own tension $T_i(r_i)\, , \quad i=
1,\ldots ,n-1$. We suppose that a boundary $\partial M^{(5)}$ also
corresponds to two hypersurfaces $r=\const : \quad r=r_0$ and
$r=r_{n}\, $ and, either 4--D geometry on $\partial M^{(5)}$ is
closed (induced 4--D metric vanishes there), or opposite points
$r_0$ and $r_{n}$ are identified with each other. In the first
case boundary terms corresponding to $\partial M^{(5)}$ are equal
to zero. In the second case, the boundary $\partial M^{(5)}$ is
absent, however, if the geometry is not smoothly matched here, it
results in appearance of an additional brane with a tension
$T_0(r_0)$). For simplicity, bulk matter is considered in the form
of a cosmological constant, in general, different for each of
$M^{(5)}_i$. Thus, our model is described by the following action:
%%%%%%
\be{2.1} S^{(5)} = \frac 1{2\kappa^2_5}\int\limits_{M^{(5)}}d^{\,
5}X\sqrt{|g^{(5)}|}\left( R[g^{(5)}] - 2 \Lambda_5 (r)\right) +
S_{YGH} -\left.\sum _{i =0}^{n-1}T_i(r_i) \int d^{\,
4}x\sqrt{|g^{(4)}|}\right|_{r_i} \, , \ee
%%%%%%
where $S_{YGH} = - \kappa^{-1}_5 \int_{\partial M^{(5)}} d^{\,
4}x\sqrt{|g^{(4)}|}\, K $ is the standard York - Gibbons - Hawking
boundary term\footnote{In compact brane-world models it is worthy
to include this term even if boundary $\partial M^{(5)}$ is absent
because it is convenient here (as well as for all models with
branes) to split manifold $M^{(5)}$ by branes into n submanifolds:
$M^{(5)} = \cup_{i=1}^n M^{(5)}_i$ each of them has boundaries
$\partial M^{(5)}_i $ defined by positions of the branes. Such
boundary terms at $\partial M^{(5)}_i $ take into account the
presence of the branes and are needed in order to satisfy the
variational principle and the junction conditions on the branes
\cite{Silva}. These junction conditions coincide with ones
following directly from the Einstein equation \rf{2.2}.\label{2}}.
The Einstein equation corresponding to action \rf{2.1} reads:
%%%%%%%
\be{2.2} R_{MN}[g^{(5)}] -\frac12 g^{(5)}_{MN} R [g^{(5)}] = -
\Lambda_5 (r) g^{(5)}_{MN} - \frac{\kappa^2_5}{\sqrt{|g^{(5)}|}}
\sum _{i =0}^{n-1}T_i(r_i)\sqrt{|g^{(4)}(x,r_i)|}\, \,
g^{(4)}_{\mu\nu}(x,r_i)
\delta_M^{\mu}\delta_N^{\nu}\delta(r-r_i)\, , \ee
%%%%%%%
In equations \rf{2.1} and \rf{2.2}
%%%%%%
\be{2.3} \Lambda_5 (r) := \sum_{i=1}^{n} \Lambda_i \theta_i (r)\,
,\quad \Lambda_i = \const \, \, , \ee
%%%%%%
with piecewise discontinuous functions
%%%%%%
\be{2.4} \theta_i (r) = \eta (r-r_{i-1} ) - \eta (r-r_i)
=\left\{\begin{array}{rcl} &0&\, , \quad r < r_{i-1}\\ &1&\, ,
\quad r_{i-1} < r < r_i \\ &0&\, , \quad r > r_i \\
\end{array}\right. \ee
%%%%%%%%
where step functions $\eta ( r-r_i)$ equal to zero for $r<r_i$ and
unity for $r>r_i$.

Now, we suppose that a metric\footnote{Different parts of the
manifold $M^{(5)}$ can be covered by different coordinates charts.
We show an explicit example below.\label{3}}
%%%%%%%%
\ba{2.5} g^{(5)}(X) &=& g^{(5)}_{MN}dX^M\otimes dX^N = dr\otimes
dr + a^2(r)\gamma^{(4)}_{\mu \nu}(x)dx^{\mu}\otimes dx^{\nu}\, \,
,
\\ a(r)& = & \sum_{i=1}^{n} a_i (r) \theta_i (r)
\nn \ea
%%%%%%%%%
is the solution of the Einstein equation \rf{2.2} and has
following matching conditions: $ a_i (r_i) = a_{i+1} (r_i) \; ,\;
i = 1,\ldots ,n-1 $ and $ a_1 (r_0) = a_n (r_n) $. Scale factors
$a_i(r)$ are supposed to be non-negative smooth functions in
intervals $[r_{i-1},r_i]$. Boundary points $r_0$ and $r_n$ are
either identified with each other: $ r_0 \leftrightarrow r_n $ or
they are not identified and the geometry in latter case is closed:
$ a_1 (r_0) = a_n (r_n) = 0\, $; i.e. induced metric
$g^{(4)}_{\mu\nu} (x,r) = a^2(r) \gamma^{(4)}_{\mu\nu} (x)$
vanishes in this points.

Having at hands solution \rf{2.5}, we can perform dimensional
reduction of action \rf{2.1}. Here, the dimensional reduction
means integration over extra dimension in 5--D part of action
\rf{2.1} to get 4--D effective action. To do that, let us perform
first some preliminary calculations.

Applying equation \rf{a5} to our case we obtain
%%%%%%%%%%%%
\be{2.6} R [g^{(5)}] = a^{-2}(r) R [\gamma^{(4)}] - f_1(r) \, ,
\ee
%%%%%%%%%%%
where
%%%%%%%%%%%
\be{2.7} f_1(r) := 8 \frac{a^{''}}{a} + 12 \left(
\frac{a^{'}}{a}\right)^2 \, . \ee
%%%%%%%%%%%
Using properties of $\theta$-function : $ \theta_i^{\, p} =
\theta_i\; ,\; p>0\; ;\quad \theta_i \, \theta_j = 0 \; ,\; i \ne
j\; \Longrightarrow a^p = \sum_{i=1}^n a_i^p \theta_i \; ,\;
\forall p$ and $ \theta_i^{'} = \delta (r-r_{i-1}) - \delta
(r-r_i)$, the function $f_i (r)$ can be written in the following
form:
%%%%%%%%%%%%
\be{2.8} f_1 (r) = 12 \sum_{i=1}^n \frac {(a^{'}_i)^2}{a_i^2}
\theta_i + 8 \sum_{i=1}^n \frac {a^{''}_i}{a_i} \theta_i - 2
\left[ K (r^+_0) \delta (r-r_0) - K (r_n^-) \delta (r-r_n) +
\sum_{i=1}^{n-1} \widehat K (r_i) \delta (r-r_i) \right]\, ,\ee
%%%%%%%%%%%
where $\widehat K (r_i) = K (r_i^+) - K (r_i^-)$ and $K (r_i^+) =
- \left. 4 a^{'}_{i+1} /a_{i+1} \right|_{r_i^{+}}\; ,\quad K
(r_i^-) = - \left. 4 a^{'}_{i} /a_{i} \right|_{r_i^{-}}$ in
accordance with equation \rf{b6}. As we can see, function $f_1$
contains all information about boundary terms and for correct
dimensional reduction of action \rf{2.1} it is not necessary to
include additionally boundary term $S_{YGH}$ because it will lead
in this case to its double counting\footnote{There are two
equivalent ways of the dimensional reduction. First, we can divide
action integral \rf{2.1} into $n$ integrals in accordance with the
splitting procedure described in footnote \ref{2} and take into
account the boundary terms at $\partial M_i^{(5)}$ arising due to
the presence of branes. In this case, the scale factors $a_i(r)$
for each of the submanifold $M_i^{(5)}$ are smooth functions and
their derivatives do not result in $\delta$--functions. Here, the
brane boundary terms are taken into account directly in the action
functional. In second approach, we consider full non-split action
\rf{2.1} without the brane boundary terms but take into account
that scale factor $a(r)$ is not a smooth function in points
corresponding to the branes location. Thus, its second derivative
has $\delta$--function terms which completely equivalent to the
brane boundary terms (see \rf{2.8}). It can be easily checked that
integration over extra dimension in both of these approaches
results in the same 4--D effective action. In the present paper we
applied second approach. \label{reduction}}. It can be easily seen
also that integral
%%%%%%%%%%%
\be{2.9} \int_{r_0}^{r_n} dr a^4 (r) f_1 (r) = - 12 \sum_{i = 1}^n
\int_{r_{i-1}}^{r_i} dr a_i^2 \, \left( a_i^{'} \right)^2 \, , \ee
%%%%%%%%%%%
where we used integration by parts. Thus, dimensional reduction of
action \rf{2.1} will result in the following effective 4--D
action:
%%%%%%%%%%
\be{2.10} S^{(4)}_{eff} = \frac
1{2\kappa^2_4}\int\limits_{M^{(4)}}d^{\,
4}x\sqrt{|\gamma^{(4)}|}\; \left\{ R[\gamma^{(4)}]\; -\; 2
\Lambda^{(4)}_{eff} \right\}\, , \ee
%%%%%%%%%%
where effective 4--D cosmological constant is
%%%%%%%%%%
\be{2.11} \Lambda^{(4)}_{eff} = \frac{1}{B_0} \left[ B_1\; +\;
B_2\; + \; B_3 \right] \ee
%%%%%%%%%%
and
%%%%%%%%%%
\ba{2.12} B_0 &=& \sum_{i=1}^n \int_{r_{i-1}}^{r_i} dr a_i^2 \; ,
\\ B_1 &=& -6\sum_{i=1}^n \int_{r_{i-1}}^{r_i} dr a_i^2
\left( a_i^{'} \right)^2\; ,\label{2.13} \\ B_2 &=& \sum_{i=1}^n
\Lambda_i \int_{r_{i-1}}^{r_i} dr a_i^4 \; , \label{2.14} \\ B_3
&=& \kappa^2_5 \sum_{i=0}^{n-1} a_i^4 (r_i) T_i (r_i) \;
..\label{2.15} \ea
%%%%%%%%%%
Effective 4--D gravitational constant is defined as
follows\footnote{In this paper we focus on the problem of
stability of considered models and not discuss cosmology on
branes. It is clear that from the point of an observer on a brane,
physical metric is the induced metric on this brane (let it be
i-th brane): $g^{(4)}_{(ph)\mu\nu} =
a_i^2(r_i)\gamma^{(4)}_{\mu\nu}$. It means we should perform
evident substitutions $a_j(r) \to a_j(r)/a_i(r_i)$ in
corresponding formulas. For example, for this observer physical
effective 4--D cosmological and gravitational constants read as
follows: $\Lambda^{(4)}_{eff} \to \Lambda^{(4)}_{(ph) eff} =
\Lambda^{(4)}_{eff} /a^2_i(r_i)$ and $\kappa^2_4 \to
\kappa^2_{(ph) 4} = \kappa^2_4 a^2_i(r_i)$. On proportionality of
the effective 4--D Newton's constant on brane to $\left.
a^2(r)\right|_{brane}$ was pointed e.g. in \cite{CGR} \label{4}.}:
$\; \kappa^2_4 = \kappa^2_5 / B_0 $\, . Equation \rf{2.10} shows
that solution \rf{2.5} of eq. \rf{2.2} takes place only if 4--D
metric $\gamma^{(4)}$ is the Einstein space metric.

\medskip

\subsection{ Examples}

In this subsection we apply considered above procedure of the
dimension reduction to some well known solutions (see e.g.
\cite{RS,CF,KKS,RSII,GS}).
\bigskip

{\bf{ a)$\;$ Poincar\' e branes }}

\medskip

In this model\footnote{Here, we follow the original solution
\cite{RS}, where scale factors are dimensionless.\label{5}} $r_0
=- L\, ,\; r_1 = 0\, ,\; r_2 = L\, $ ,
%%%%%%%%%%
\be{2.16} \vphantom{\int} \begin{array}{rcl} a_1 (r) &=& \exp (r /
l)\; ,\phantom{\frac{R^l_i}{d^l_i}}\, \,
\quad -L \le r \le 0\,
,\\ a_2 (r) &=& \exp (-r / l)\; , \phantom{\frac{R^l_i}{d^l_i}}
\qquad 0 \le r \le L\, ,\\
\end{array}
\ee
%%%%%%%%%%%
and bulk cosmological constants $\Lambda_1 = \Lambda_2 = -6 /
l^2$, where $l$ is the AdS radius. The points $r_0$ and $r_2$ are
identified with each other. A free parameter $L$ defines the size
of models in the additional dimension. The geometry is not smooth
at points $r=0$ and $r=r_0 \equiv r_2$, thus we have two branes
with tensions: $-T_0(r_0)= T_1(r_1) = 6/\left(\kappa^2_5\,
l\right)$. Substituting concrete expressions into formulas
\rf{2.12} - \rf{2.15}, we obtain respectively:
%%%%%%%%%%
\ba{2.17} B_0 &=& l \left( 1 - e^{-2L / l} \right) > 0 \; ,\\ B_1
&=& \frac3l \left( e^{-4L / l} -1 \right)\label{2.18} \; ,
\\ B_2 &=& \frac3l \left( e^{-4L / l} -1 \right)
\; ,\label{2.19} \\ B_3 &=& \frac6l \left( 1 - e^{-4L / l}
\right)\; . \label{2.20}\ea
%%%%%%%%%%
Thus, in this model
%%%%%%%%%%%
\be{2.21} \Lambda^{(4)}_{eff} \equiv 0 \ee
%%%%%%%%%%%
and $\gamma^{(4)}_{\mu\nu}$ is flat space-time metric. The
Randall--Sundrum one brane solution \cite{RSII} corresponds to
trivial limite $L \longrightarrow + \infty$ and also results in
eq. \rf{2.21}. In this case, the extra coordinate $r$ runs over
$\RR$, but all integrals of the type \rf{2.12} - \rf{2.15} are
convergent due to exponential decrease of the warp factors
$a_{1,2}$ when $|r| \to \infty$. Effectively, this model is
compact with respect to the extra dimension.

\medskip

{\bf{ b)$\;$ De Sitter brane (symmetric solution) }}

\medskip

In this model $r_0 = 0\, ,\; r_1 = L\, ,\; r_2 = 2L\, $ ,
%%%%%%%%%%
\be{2.22} \vphantom{\int} \begin{array}{rcl} a_1 (r) &=& l\, \sinh
\frac{r}{l}\; ,\phantom{\frac{R^l_i}{d^l_i}}\; \;\qquad 0 \le r
\le L\, ,\\ a_2 (r) &=& l\, \sinh \frac{2L - r }{l}\; ,
\phantom{\frac{R^l_i}{d^l_i}} \quad L\le r \le 2L\, ,\\
\end{array}
\ee
%%%%%%%%%%%
and bulk cosmological constants $\Lambda_1 = \Lambda_2 = -6 /
l^2\, $. In the points $r_0$ and $r_2$ the geometry is closed:
$a_1 (r_0) = a_2 (r_2) = 0$ ($r_0$ and $r_2$ are horizons of
$AdS_5$). The geometry is not smooth in $r_1$. Therefore, in this
model we have one brane with tensions: $ T_1(r_1) =\left[
6/\left(\kappa^2_5\, l\right)\right] \coth\left(L / l\right)$.
%This one-brane solution was described in detail in paper \cite{GS}.
Substituting these expressions into
formulas \rf{2.12} - \rf{2.15}, we obtain respectively:
%%%%%%%%%%
\ba{2.23} B_0 &=& l^3 \left(\frac12 \sinh \frac{2L}{l} -
\frac{L}{l} \right) > 0\; ,\\ B_1 &=& -3 l^3 \left( \sinh^3
\frac{L}{l} \cosh \frac{L}{l} + \frac14 \sinh 2\frac{L}{l}
-\frac{1}{2} \frac{L}{l} \right)\label{2.24} \; ,
\\ B_2 &=& -3 l^3 \left( \sinh^3
\frac{L}{l} \cosh \frac{L}{l} -\frac{3}{4} \sinh 2\frac{L}{l} +
\frac{3}{2} \frac{L}{l} \right) \; ,\label{2.25} \\ B_3 &=& 6 l^3
\sinh^3 \frac{L}{l} \cosh \frac{L}{l}\; . \label{2.26}\ea
%%%%%%%%%%
So, in this model
%%%%%%%%%%%
\be{2.27} \Lambda^{(4)}_{eff} = 3 \ee
%%%%%%%%%%%
and $\gamma^{(4)}_{\mu\nu}$ describes either Riemannian 4--sphere
with scalar curvature $R [\gamma^{(4)}] = D_0 \left( D_0 -
1\right) = 12 $ or 4--D de Sitter space-time with cosmological
constant $\Lambda = 3$ and scalar curvature $R [\gamma^{(4)}] =
\left[ 2 D_0 /( D_0 - 2)\right]\Lambda = 12 $.

\medskip

{\bf{ c)$\;$ De Sitter brane (non-symmetric solution)}}

\medskip

We obtain this solution gluing together two submanifolds covered
by different charts. First submanifold describes truncated
Garriga--Sasaki instanton \cite{GS} and second one describes flat
5--D space:
%In this model $r_0 = 0\, ,\; r_1 = L\, ,\; r_2 = 2L\, $ ,
%%%%%%%%%%
\be{2.28} \vphantom{\int} \begin{array}{rcl} a_1 (r) &=& l\, \sinh
\frac{r}{l}\; ,\phantom{\frac{R^l_i}{d^l_i}}\; \quad 0 \le r \le
L\, ,\\ a_2 (R) &=& R_0 - R \; , \phantom{\frac{R^l_i}{d^l_i}}
\quad 0\le R \le R_0\, ,\\
\end{array}
\ee
%%%%%%%%%%%
where $R_0 = l\, \sinh (L / l)$. Bulk cosmological constants
$\Lambda_1 = -6 / l^2\, $ and $\Lambda_2 \equiv 0$. In the points
$r = 0$ and $R = R_0$ the geometry is closed: $a_1 (0) = a_2 (R_0)
= 0$. The geometry is not smooth on the hypersurfaces of gluing $
r = L$ and $R=0$. That why, we have one brane with tensions: $
\left. T_1 \right|_{r=L\, , R=0} = 3/ \kappa^2_5 \left[ 1/R_0 +
(1/l) \coth\left(L / l\right) \right]$. Substituting these
expressions into formulas \rf{2.12} - \rf{2.15}, we obtain
respectively:
%%%%%%%%%%
\ba{2.29} B_0 &=& \frac12 l^3 \left(\frac12 \sinh \frac{2L}{l} -
\frac{L}{l} \right) + \frac13 R_0^3 > 0\; ,\\ B_1 &=& -\frac32 l^3
\left( \sinh^3 \frac{L}{l} \cosh \frac{L}{l} + \frac14 \sinh
2\frac{L}{l} -\frac{1}{2} \frac{L}{l} \right) -2 R_0^3
\label{2.30} \; ,
\\ B_2 &=& -\frac32 l^3 \left( \sinh^3
\frac{L}{l} \cosh \frac{L}{l} -\frac{3}{4} \sinh 2\frac{L}{l} +
\frac{3}{2} \frac{L}{l} \right) \; ,\label{2.31} \\ B_3 &=& 3 l^3
\sinh^3 \frac{L}{l} \cosh \frac{L}{l} + 3 R_0^3\; .
\label{2.32}\ea
%%%%%%%%%%
Therefore, as in the symmetric case, in this model
%%%%%%%%%%%
\be{2.33} \Lambda^{(4)}_{eff} = 3 \ee
%%%%%%%%%%%
and $\gamma^{(4)}_{\mu\nu}$ describes either 4--sphere or the de
Sitter space with cosmological constant $\Lambda = 3$.

\medskip

{\bf{ d)$\;$ Anti de Sitter brane }}

\medskip

In this model $r_0 = -L\, ,\; r_1 = 0\, ,\; r_2 = L\, $ ,
%%%%%%%%%%
\be{2.34} \vphantom{\int} \begin{array}{rcl} a_1 (r) &=& l\, \cosh
\frac{L+r}{l}\; ,\phantom{\frac{R^l_i}{d^l_i}}\; \; -L\le r \le
0\, ,\\ a_2 (r) &=& l\, \cosh \frac{L - r }{l}\; ,
\phantom{\frac{R^l_i}{d^l_i}} \qquad 0\le r \le L\, ,\\
\end{array}
\ee
%%%%%%%%%%%
and bulk cosmological constants $\Lambda_1 = \Lambda_2 = -6 /
l^2\, $. The points $r_0$ and $r_2$ are identified with each
other. The geometry is not closed here and can be smoothly glued
in this points. The points $r_{0,2}$ correspond to wormhole
throats in the Riemannien space. The geometry is not smooth in
$r_1$. Therefore, in this model we have only one
brane\footnote{This model can be easily generalized to the case of
an arbitrary number of parallel branes by gluing one-brane
manifolds at throats and identifying the two final opposite
throats.\label{6}} with tension: $ T_1(r_1) =\left[
6/\left(\kappa^2_5\, l\right)\right] \tanh\left(L / l\right)$.
%This one-brane solution was described in detail in paper \cite{GS}.
Substituting these expressions into formulas \rf{2.12} -
\rf{2.15}, we obtain respectively:
%%%%%%%%%%
\ba{2.35} B_0 &=& l^3 \left(\frac12 \sinh \frac{2L}{l}
+\frac{L}{l} \right) > 0\; ,\\ B_1 &=& -3 l^3 \left( \sinh^3
\frac{L}{l} \cosh \frac{L}{l} + \frac14 \sinh 2\frac{L}{l}
-\frac{1}{2} \frac{L}{l} \right)\label{2.36} \; ,
\\ B_2 &=& -3 l^3 \left( \sinh^3
\frac{L}{l} \cosh \frac{L}{l} +\frac{5}{4} \sinh 2\frac{L}{l} +
\frac{3}{2} \frac{L}{l} \right) \; ,\label{2.37} \\ B_3 &=& 6 l^3
\left( \sinh^3 \frac{L}{l} \cosh \frac{L}{l} + \frac12 \sinh 2
\frac{L}{l}\right)\; . \label{2.38}\ea
%%%%%%%%%%
Thus, in this model
%%%%%%%%%%%
\be{2.39} \Lambda^{(4)}_{eff} = - 3 \ee
%%%%%%%%%%%
and $\gamma^{(4)}_{\mu\nu}$ describes either Riemannian
4--hyperboloid with scalar curvature $R [\gamma^{(4)}] = - D_0
\left( D_0 - 1\right) = - 12 $ or 4--D Anti de Sitter space-time
with cosmological constant $\Lambda = - 3$ and scalar curvature $R
[\gamma^{(4)}] = \left[ 2 D_0 /( D_0 - 2)\right]\Lambda = -12 $.

\medskip

To conclude this section, we consider in more details the Einstein
equation \rf{2.2} with the help of formulas \rf{a1} - \rf{a5}. In
addition to equation \rf{2.6} we obtain:
%%%%%%%%%%%%
\ba{2.40} R_{rr} [ g^{(5)} ] &=& -4\, \frac{a^{''}}{a}\, , \nn
\\ R_{\mu r}[ g^{(5)}] &=& R_{r \mu}
[g^{(5)}] = 0\, , \\ R_{\mu\nu}[g^{(5)}] &=&
R_{\mu\nu}[\gamma^{(4)}] - a^2
\gamma^{(4)}_{\mu\nu}\left[\frac{a^{''}}{a} +
3\left(\frac{a^{'}}{a}\right)^2 \right]\, .\nn \ea
%%%%%%%%%%%%
Then, $rr$ and $\mu \nu$-components of eq. \rf{2.2} are reduced
correspondingly to equations:
%%%%%%%%%%
\be{2.41} R[\gamma^{(4)}] =2 \sum_{i=1}^n a_i^2 (r) \Lambda_i
\theta_i (r)+ 12 \sum_{i=1}^n \left( a_i^{'}\right)^2 \theta_i
(r)\equiv f_2(r)\, \ee
%%%%%%%%%%%
and
%%%%%%%%%%
\be{2.42} R_{\mu\nu}[\gamma^{(4)}] - \frac12 \gamma^{(4)}_{\mu\nu}
R [\gamma^{(4)}] = - \gamma^{(4)}_{\mu\nu} \left\{ 3\sum_{i=1}^n
\left(a_i^{'}\right)^2\theta_i + \sum_{i=1}^n \Lambda_i
a_i^2\theta_i +3 \sum_{i=1}^n a_i a_i^{''}\theta_i\right\} \equiv
- \gamma^{(4)}_{\mu\nu} f_3 (r) \, .\ee
%%%%%%%%%%
In latter equation $\delta$-function terms originated from
$a^{''}$ and tension terms cancel each other. It can be easily
seen that for the Poincar\' e brane model we obtain $f_2(r) \equiv
f_3(r) \equiv 0$ in accordance with eq. \rf{2.21}. For symmetric
de Sitter brane model $f_2(r) \equiv 12$ and $f_3(r) \equiv 3$,
which corresponds to eq. \rf{2.27}. In non-symmetric de Sitter
brane model: $f_2(r) \equiv f_2(R) \equiv 12$ and $f_3(r) \equiv
f_3(R) \equiv 3$ in accordance with eq. \rf{2.33}. For the Anti de
Sitter brane model $f_2(r) \equiv -12$ and $f_3(r) \equiv -3$,
which corresponds to eq. \rf{2.39}.

%%%%%%%%%%%%%%%%%%%%%%%%%%%%%%%%%%%%%%%%%%%%%%%%%%%%%%%%%%%%%%
\section{Stability under conformal excitations\label{stability}}
\setcounter{equation}{0}

\bigskip

Let us investigate now the stability of metric $g^{(5)}(X)$
defined in \rf{2.5} with respect to conformal excitations. In
other words, we want to investigate the dynamical behaviour of the
conformal metric excitations developing on the fixed background
$g^{(5)} (X)$.
%which satisfies equation \rf{2.2}.
To do this, we consider a perturbed metric of the form of \rf{b1}:
$\bar g^{(5)} = \Omega^2 g^{(5)} \equiv e^{2\beta } g^{(5)}$,
where $\Omega =1$ corresponds to the background solution and
$\beta \ll 1$ describes the small perturbation limit. Obviously,
the background solution is stable against such perturbations if
$\Omega$ oscillates with time near the value $\Omega =1$ and is
unstable if $\Omega$ has a runaway behaviour from this value.
According to the Perturbation Theory, full analysis should consist
of two steps. The first one is the investigation of the dynamical
behaviour of perturbations on the fixed background, and the second
one is the study of the back reaction of perturbations on the
background solution. In the present paper we are concentrating on
the first problem, to find which brane-world solutions are stable
against the conformal perturbations, and postpone the second
problem for our future investigation.

According to the standard approach, equation of motion for
perturbations (in our case for $\Omega$ or $\beta$), developing on
the fixed background $g^{(5)}$, can be obtained substituting $\bar
g^{(5)}$ in equation \rf{2.2} and taking into account the
background solution $g^{(5)}$ (e.g. solutions from section 2.1).
However, it is possible to investigate this problem by a different
way: starting from action \rf{2.1}, putting there perturbed metric
$\bar g^{(5)}$ and, after that, taking into account the background
solutions $g^{(5)}$. Then, the resulting effective action will
describe the dynamical behaviour of the perturbations on the fixed
background. From this action we can obtain the energy momentum
tensor of the perturbations to study the back reaction of them on
the background metric.

In our paper we follow the second approach and put the perturbed
metric $\bar g^{(5)}$ into action \rf{2.1} which
yields\footnote{It is clear that for conformally transformed
metric the Lanczos-Israel junction conditions will change (see
e.g. eq. \rf{b8}). But, at the moment, we do not consider a back
reaction of the conformal excitations on the metric; i.e. on the
behaviour of $a(r)$, and on the junction condition.\label{7}}.:
%%%%%%%%%%%
\be{3.1} \bar S^{(5)} = \frac
1{2\kappa^2_5}\int\limits_{M^{(5)}}d^{\, 5}X\sqrt{|\bar
g^{(5)}|}\left( R[\bar g^{(5)}] - 2 \Lambda_5 (r)\right)
-\left.\sum _{i =0}^{n-1}T_i(r_i) \int d^{\, 4}x\sqrt{|\bar
g^{(4)}|}\right|_{r_i} \, ,\ee
%%%%%%%%%%%
%where we suppose that matter is keeping without changes. It means
%that the bulk cosmological constant as well as the tensions of
%branes ("vacuum energies" on branes) do not change\footnote{It is
%clear that for conformally transformed metric the Lanczos-Israel
%junction conditions will change (see e.g. eq. \rf{b8}). But, at
%the moment, we do not consider a back reaction of the conformal
%excitations on the metric; i.e. on the behaviour of $a(r)$, and on
%the junction condition.\label{7}}.
With the help of equations \rf{b3} and \rf{2.6} the first term in
this action reads:
%%%%%%%%%%%
\ba{3.2} \sqrt{|\bar g^{(5)}|}\, R[\bar g^{(5)}] &=& \Omega^5
a^4\; \sqrt{|\gamma^{(4)}|} \left\{ \right. \Omega^{-2} \left[
a^{-2} R[\gamma^{(4)}] - f_1(r) \right] - \\ &-&
8\Omega^{-3}\Omega_{;M;N} g^{(5)MN} - 4
\Omega^{-4}\Omega_{,M}\Omega_{,N}g^{(5)MN } \left. \right\} \,
.\nn \ea
%%%%%%%%%%%
For generality, we do not assumed the small perturbation limit
$\beta \ll 1$ keeping in action all non-linear perturbation terms.
Transition to this limit can be easily performed in the final
expression (see equation \rf{3.12} below). In what follows, we
shall consider a particular case when conformal prefactor is a
function of 4--D space-time coordinates: $\Omega = \Omega (x)
\equiv \exp (\beta (x) )$. It is well known that conformal
excitations of this form behaves as scalar fields in 4--D
space-time (e.g. on branes ). Because of the prefactor $\Omega^3
(x)$ in front of 4--D scalar curvature $R[\gamma^{(4)}]$ in
action, the 4--D metric $\gamma^{(4)}$ is written in the
Brans--Dicke frame. However, it is more easy to investigate the
conformal perturbation stability in the Einstein frame (in the
introduction we mentioned the equivalence of these frames with
respect to the stability analysis) :
%%%%%%%%%%
\be{3.3} \gamma^{(4)}_{\mu\nu} (x) \Longrightarrow \tilde
\gamma^{(4)}_{\mu\nu} (x) = \Omega^3 (x)\gamma^{(4)}_{\mu\nu}
(x)\, . \ee
%%%%%%%%%
In this frame dimensionally reduced action \rf{3.1} reads
%%%%%%%%%%%
\be{3.4} \bar S^{(4)}_{eff} = \frac
1{2\kappa^2_4}\int\limits_{M^{(4)}}d^{\, 4}x\sqrt{|\tilde
\gamma^{(4)}|}\; R[\tilde \gamma^{(4)}]\; + \; \frac
12\int\limits_{M^{(4)}}d^{\, 4}x\sqrt{|\tilde
\gamma^{(4)}|}\;\left( - \tilde \gamma^{(4)\mu\nu} \tilde
\beta_{,\mu} \tilde \beta_{,\nu}\; -\; 2 \widetilde U_{eff}
\right)\, , \ee
%%%%%%%%%%%
where $\tilde \beta \equiv \sqrt{3/2} (1/\kappa_4) \beta $ and
%%%%%%%%%%%
\be{3.5} \widetilde U_{eff} (\Omega ) \equiv
\frac{1}{\kappa_4^{2}} U_{eff} (\Omega ) = \frac{1}{\kappa_4^2
B_0} \left[ B_1\Omega^{-3}\; +\; B_2\Omega^{-1}\; + \;
B_3\Omega^{-2} \right]\, .\ee
%%%%%%%%%%
Here, parameters $B_i \; (i=0,\ldots ,3)$ are defined by equations
\rf{2.12} - \rf{2.15}.

Now, the problem of the background solution \rf{2.5} stability
against the conformal excitations is reduced to existence of a
minimum of the effective potential $U_{eff}$ at point $\Omega = 1
\Longleftrightarrow \beta = 0$ which corresponds to the absence of
the perturbations. In Appendix B we show additionally that all
other values for the minimum lead to metrics which in zero order
do not satisfy the same Einstein equation as the background
solution \rf{2.5}. It is clear also that the effective
cosmological constant \rf{2.11} should coincide with $U_{eff}$ at
$\Omega = 1\, : \; \Lambda_{eff}^{(4)} = U_{eff} (\Omega = 1)$
which we explicitly obtain from \rf{3.5}.
%natural in zero order approximation ( where
%considered system is in position of minimum and conformal
%excitations are absent ).
The extremum existence condition reads:
%%%%%%%%%%
\be{3.6} \left. \frac{\partial U_{eff}}{\partial \Omega}
\right|_{\Omega = 1} = 0 \Longrightarrow 3B_1 + B_2 + 2B_3 = 0\, .
\ee
%%%%%%%%%
Small excitations near a minimum position can be observed on
branes as massive scalar fields - gravitational excitons with mass
squared:
%%%%%%%%%
\be{3.7} m^2 = \left. \frac{\partial^2 \widetilde
U_{eff}}{\partial \tilde \beta^2}\right|_{\tilde \beta = 0} =
\left. \frac23 \Omega^2 \frac{\partial^2 U_{eff}}{\partial
\Omega^2} \right|_{\Omega = 1} = \frac{2}{3B_0} \left( 12 B_1 + 2
B_2 + 6 B_3 \right)\, .\ee
%%%%%%%%%
Obviously, the original solution \rf{2.5} is stable under these
conformal excitations iff $m^2 > 0$ which prevent their runaway
behaviour from the background solution. As it can be easily seen,
all four models considered in previous section satisfy equation
\rf{3.6}. It means that all these solutions are stationary points
of $U_{eff}$ if the effective potential is considered as a
functional of $a(r)$.
%that would be natural to expect.
For masses squared in the case of the Poincar\' e, the de Sitter
(symmetric solution), the de Sitter (non-symmetric solution) and
the Anti de Sitter branes we obtain respectively:
%%%%%%%%%%%
\ba{3.8} m^2 &=& \frac{4}{l B_0} \left( e^{-4L/l} -1\right) \, <
\, 0 \, , \\ m^2 &=& \frac{4l^3}{B_0} \left( - \sinh^3 \frac{L}{l}
\cosh \frac{L}{l} - \frac34 \sinh 2 \frac{L}{l} + \frac32
\frac{L}{l} \right) \, < \, 0 \, ,\label{3.9}\\ m^2 &=&
\frac{2l^3}{B_0} \left( - \sinh^3 \frac{L}{l} \cosh \frac{L}{l} -
\frac34 \sinh 2 \frac{L}{l} + \frac32 \frac{L}{l} -
2\frac{R_0^3}{l^3} \right) \, < \, 0 \, ,\label{3.10}\\ m^2 &=&
\frac{4l^3}{B_0} \left( - \sinh^3 \frac{L}{l} \cosh \frac{L}{l} +
\frac14 \sinh 2 \frac{L}{l} + \frac32 \frac{L}{l} \right) \,
\lesseqqgtr \, 0 \, .\label{3.11} \ea
%%%%%%%%%%%%
Thus, three first solutions are unstable under considered
conformal excitations: $\Omega = 1$ corresponds to maximum but not
to minimum of the potential \rf{3.5}. However, in the AdS brane
case, mass squared is positive and decreases from\footnote{Masses
squared of gravexcitons \rf{3.8} - \rf{3.11} are written in
dimensionless units. If we take into account footnote \ref{4},
then physical gravexciton mass for an observer on i-th brane is:
$m \to m_{(ph)} = m/a_i(r_i)$.\label{8}} 4 for $L/l \to 0$ to zero
for $L/l \to 1$ (more precisely, numerical calculations show that
$m^2 \to 0$ for $L/l \to 0,988$). So, the AdS brane solution is
stable with respect to the conformal excitations if the distance
between brane and throats of wormholes is less than the AdS
radius. As it was mentioned in the footnote \ref{6}, this case can
be easily generalized to a number of AdS branes connected with
each other via the wormhole throats. Then, in the case of $n$
branes, for the gravexciton mass squared we obtain an expression
which is simply an algebraic sum of the type \rf{3.11} equations
(with an evident substitution $L \to L_i \, , \; i = 1,\ldots ,n$
for each member of the sum) and overall prefactor $B_0^{-1}$ .
Here, $B_0$ is also a generalization of \rf{2.35} to an evident
sum. This mass squared is positive e.g. if $L_i /l \lesssim 1\, ,
\; i = 1,\ldots ,n$.

For small fluctuations near minimum of $U_{eff}$ action \rf{3.4}
reads:
%%%%%%%%%%%%%%%%%
\be{3.12} \bar S^{(4)}_{eff} = \frac
1{2\kappa^2_4}\int\limits_{M^{(4)}}d^{\, 4}x\sqrt{|\tilde
\gamma^{(4)}|}\; \left\{ R[\tilde \gamma^{(4)}] -
2\Lambda^{(4)}_{eff} \right\}\; + \; \frac
12\int\limits_{M^{(4)}}d^{\, 4}x\sqrt{|\tilde
\gamma^{(4)}|}\;\left( - \tilde \gamma^{\mu\nu} \tilde
\beta_{,\mu} \tilde \beta_{,\nu}\; -\; m^2\tilde\beta^2\right)\, ,
\ee
%%%%%%%%%%%
where the first integral corresponds to zero order theory
\rf{2.10} (background solution) and the second one describes
gravitational excitons. This effective action can be used for
investigation of the gravexciton back reaction on the background
metric.

If we put in action \rf{3.1} conformally transformed brane
tensions $\bar T (r_i) = (1 /\Omega) T(r_i)$ (see \rf{b7} and
\rf{b8}) instead of $T(r_i)$, the effective potential reads:
%%%%%%%%%%%
\be{3.13} U_{eff} (\Omega ) = \frac{1}{B_0} \left[
B_1\Omega^{-3}\; +\; B_2\Omega^{-1}\; + \; B_3\Omega^{-3}
\right]\, .\ee
%%%%%%%%%%
In this case, $\Omega = 1$ is not the extremum of the effective
potential \rf{3.13} for all of four considered solutions: they are
not stationary points of this potential.
%%%%%%%%%%%%%%%%%%%%%%%%%%%%%

%%%%%%%%%%%%%%%%%%%%%%%%%%%%%%%%%%%%%%%%%%%%%%%%%%%%%%%%%%%%%%
\section{Conclusions\label{coclusion}}
\setcounter{equation}{0}

\bigskip

In the present paper we investigated the stability of 5--D
brane-world solutions against conformal perturbations. For these
models the 5--dimensional metric contains a 4--dimensional metric
components multiplied by a warp (scale) factor $a(r)$ which is a
function of the additional dimension. Models contain $n$ parallel
branes "transversal" to the additional coordinate. As a matter we
consider bulk cosmological constants between branes and tensions
("vacuum energies") on the branes. Scale factor is continuous
piecewise function while its derivative has jumps on the branes.
There are a number of well known exact solutions which belong to
this class of models (e.g.
\cite{RS,CF,KKS,RSII,GS,Kalop,Kim,Kogan}). We investigated
stability of some of these models under the conformal excitations
which are functions of 4--D space-time. Such excitations are of
special interest because they behave as massive minimal scalar
fields in 4--D space-time, for example can be observed as massive
scalar particles - gravitational excitons on branes,
%\footnote{Homogeneous
%conformal excitations which depend on time only are a particular
%case of gravitational excitons. They behave as a condensat on
%branes.}
as it takes place in the Kaluza-Klein approach \cite{GZ,GZII}.

To perform the stability analysis, we put the perturbed metric in
the original 5--D action, took into account the background metric
solution and integrated the action over the extra dimension (the
dimension reduction). The obtained 4--D effective action describes
the dynamical behaviour of the perturbations on the fixed
background. The extremum of the effective potential in this action
corresponds to the background solution. If this extremum is a
minimum, the conformal perturbations oscillate around the
background solution providing its stability. In this case, small
excitations around this minimum are observed as gravitational
excitons on branes. However, if this extremum is a maximum, the
conformal perturbations have the runaway behaviour, and the
background solution is unstable against such excitations.
%We performed dimensional reduction of considered model to the
%effective 4--D theories. After such reduction, dynamical behaviour
%of the conformal excitations is defined by the form of the
%effective potential. Obviously, considered models are stable under
%these fluctuations iff the effective potential has minima at
%points corresponding to the original solutions. Small excitations
%around these minima are observed as gravitational excitons on
%branes.

We have shown that in the case of one and two Poincar\' e branes,
one de Sitter brane (symmetric solution) and one de Sitter brane
(non-symmetric solution), all these solutions are unstable with
respect to these excitations because the effective potential has a
maximum but not a minimum at the point corresponding to the
original (background) solutions. In these models 4--D effective
cosmological constant is non-negative (see \rf{2.21}, \rf{2.27}
and \rf{2.33}). However, one AdS brane solution is stable if the
distance between brane and throats of wormholes is less than the
AdS radius. The effective 4--D cosmological constant is negative
in this model (see \rf{2.39}). The latter case is easily
generalized to a stable model with a number of parallel AdS branes
connected with each other via the wormhole throats.
%as we already mentioned in the introduction,
It is necessary to note that we found stability/instability
against gravexcitons for models with the same kind of branes,
which are correspondingly stable/unstable against radions
considered in papers \cite{BDL,CF}, although the direct comparison
between models can be only made for cases with two and more
branes.

%devoted to the radion stabilization
%analysis. That shows close connection between gravexcitons and
%radions for the brane-world models.

Another remark consists in the analogy between the stability under
conformal perturbations in the brane-world models considered here
and the standard Kaluza-Klein models. The situation with these
four solutions is similar to one we have in pure geometrical case
in the standard Kaluza-Klein approach \cite{GZ}. Here, the
stability also takes place when 4--D effective cosmological
constant is negative. If the effective cosmological constant is
positive we have maximum of the effective potential instead of
minimum. To shift the minimum of the effective potential to
positive values, we should include matter into the model.

We found also that 4--D effective cosmological and gravitational
constants on branes as well as gravexciton masses undergo
hierarchy. It was shown that for observers on different branes
with different warp factors these parameters have different
values. Similar result with respect to the effective 4--D Newton's
constant was obtained in \cite{CGR}.

There are a number of possible generalizations which are worth to
investigate. First, it is of interest to include more rich types
of matter in the model, e.g. perfect fluid in bulk as well as on
branes, which simulates different forms of matter in the Universe.
The presence of matter can stabilize radions in the brane-world
models with non-negative 4--D effective cosmological constant on
branes, as it was shown in \cite{CGRT}, \cite{GW}--\cite{HKP}. As
we mentioned above, stabilization of gravexcitons in models with
4--D positive effective cosmological constant takes place also in
the standard Kaluza-Klein approach if we include matter here
\cite{GZ}.
%(by analogy with the standard
%Kaluza-Klein approach \cite{GZ} as it was mentioned above).
Thus, we expect similar stabilization effect for gravexcitons in
the brane-world models. Second possibility consists in
generalization of the model to a multidimensional case with $D
> 5$. It will give opportunity to include into consideration
already obtained exact brane-world solutions with $D = 6$ and more
dimensions. Additionally, as we wrote in section 3, the
investigation of the background solution stability against the
conformal perturbations is only the first problem. When we found
the stable solutions, the second problem consists in the
investigation of the perturbation back reaction on the background
solution. We leave these issues for the future work.

%%%%%%%%%%%%%%%%%%%%%%%%%%%%%%%%%%%%%%%%%%%%%%%%%%%%%%%%%%%%%%%%

\bigskip
{\bf Acknowledgments}

We would like to thank Pedro Gonz\'alez--D{\'\i}az for stimulating
discussions and Uwe G\" unther for useful comments. A.Z. thanks
Instituto de Matem\'aticas y F\'{\i}sica Fundamental, CSIC, for kind
hospitality during preparation of this paper. A.Z. acknowledges
support by Spanish Ministry of Education, Culture and Sport (the
programme for Sabbatical Stay in Spain) and the programme SCOPES
(Scientific co-operation between Eastern Europe and Switzerland)
of the Swiss National Science Foundation, project No. 7SUPJ062239.
M.B.L. is supported by the Spanish Ministry of Science and
Technology under Research project No. PB97-1218.

%%%%%%%%%%%%%%%%%%%%%%%%%%%%%%%%%%%%%%%%%%%%%%%%%%%%%%%%%%%%%%%%

\appendix
\begin{appendix}
\section{Block-diagonal metrics \label{appendix a}}
\setcounter{equation}{0}

\bigskip

In this appendix we present some useful formulas (see also
\cite{Ivash}) for curvature tensors in the case of a
block-diagonal metric of the form:
%%%%%%%%%%
%\be{a1} g^{(D_1+D_0)}_{MN}(X)dX^M\otimes dX^N = g^{(D_1)}_{mn}(y)
%dy^m\otimes dy^n + e^{2\sigma
%(y)}g^{(D_0)}_{\mu\nu}(x)dx^{\mu}\otimes dx^{\nu} \ee
%%%%%%%%%
\be{a1} \left(\quad g^{(D_0+D_1)}_{MN}(x,y)\quad \right) =
\left(\begin{array}{cc} e^{2\sigma (y)}
\gamma^{(D_0)}_{\mu\nu}(x)&0
\\ 0 & g^{(D_1)}_{mn}(y)\end{array}\right) \, \, .\ee
%%%%%%%%%
For this metric the Ricci tensor (everywhere in this paper we use
the Misner-Thorne-Wheeler book conventions \cite{MTW}) reads
%%%%%%%%
\ba{a2} R_{\mu\nu} [ g^{(D)}] &=& R_{\mu\nu} [ \gamma^{(D_0)}] -
e^{2\sigma} \gamma^{(D_0)}_{\mu\nu}\left[ D_0\, g^{(D_1)mn}\left(
\partial_m \sigma \right)\left(\partial_n \sigma \right) +
g^{(D_1)mn}\, \nabla^{(D_1)}_m \left( \partial_n \sigma
\right)\right]\, ,\nn \\
%%%%%%%%%%%
%\be{a3}
R_{\mu n} [ g^{(D)}] &=& R_{n \mu} [ g^{(D)}] = 0\, \, ,
\\
%%%%%%%%%%%
%\be{a4}
R_{mn} [ g^{(D)}] &=& R_{mn}[ g^{(D_1)}] - D_0 \left[ \left(
\partial_m \sigma \right)\left(\partial_n \sigma \right) +
\nabla^{(D_1)}_m \left( \partial_n \sigma \right)\right]\, ,\nn
\ea
%%%%%%%%%
where $D = D_0+D_1 $ and $\nabla^{(D_1)}_m$ is a covariant
derivative with respect to the metric $g^{(D_1)}$. The scalar
curvature reads correspondingly:
%%%%%%%%%%%
\ba{a5} R [ g^{(D)}] &=& e^{- 2\sigma} R [ \gamma^{(D_0)}] + R [
g^{(D_1)}] - \nn \\ &-& D_0 \left[ \left( D_0+1
\right)g^{(D_1)mn}\left( \partial_m \sigma \right)\left(\partial_n
\sigma \right) + 2\, g^{(D_1)mn}\, \nabla^{(D_1)}_m \, \left(
\partial_n \sigma \right)\right]\, . \ea
%%%%%%%%%%%
%\end{appendix}
%%%%%%%%%%%%%%%%%%%%%%%%%%%%%%%%%%%%%%%%%%%%%%%%%%%%%%%%%%%%%%%%%%%%
%\begin{appendix}
\bigskip

\section{Conformal transformation  \label{appendix b}}
\setcounter{equation}{0}

\bigskip

For conformally transformed metric
%%%%%%%
\be{b1} \bar g^{(D)}_{MN}(X)\, = \, \Omega^2 (X)\, \, \,
g^{(D)}_{MN}(X)\, \equiv\, e^{2\beta (X)}\, \, g^{(D)}_{MN}(X) \ee
%%%%%%%%
the Ricci tensor and the scalar curvature read correspondingly:
%%%%%%%%%%
\ba{b2} R_{MN} [ \bar g^{(D)}] &=& R_{MN} [ g^{(D)}] - (D-2)
\beta_{;M;N} - g^{(D)}_{MN}\, \, g^{(D)KL} \beta_{;K;L} + \nn \\
&+& (D-2)\beta_{;M}\beta_{;N} - (D-2)\, \, g^{(D)}_{MN}\, \,
g^{(D)KL} \beta_{;K}\beta_{;L} \ea
%%%%%%%%%%%%%
and
%%%%%%%%%%%%%
\be{b3} R [ \bar g^{(D)}] = \Omega^{-2} R [ g^{(D)}] -
2(D-1)\Omega^{-3} \Omega_{;M;N}\, g^{(D)MN} -
(D-1)(D-4)\Omega^{-4} \Omega_{;M}\Omega_{;N}\, g^{(D)MN}\, ,\ee
%%%%%%%%%%%%%
where in equations \rf{b2} and \rf{b3} covariant derivatives are
taken with respect to the metric $g^{(D)}$.

We suppose now that the metric $\bar g^{(D)}$ is a solution of the
Einstein equation
%%%%%%%%%%%%%%
\ba{b4}&{}& R_{MN}[\bar g^{(D)}] -\frac12 \bar g^{(D)}_{MN} R
[\bar g^{(D)}] = \\ &-& \bar \Lambda_D \bar g^{(D)}_{MN} -
\frac{\kappa^2_D}{\sqrt{|\bar g^{(D)}|}} \sum _{i =0}^{n-1}\bar
T_i(y_i)\sqrt{|\bar g^{(D_0)}(x,y_i)|}\, \, \bar
g^{(D_0)}_{\mu\nu}(x,y_i)
\delta_M^{\mu}\delta_N^{\nu}\delta(y-y_i)\, ,\nn \ea
%%%%%%%%%%%%
which describe a model with the bulk cosmological constant $\bar
\Lambda_D$ and $n$ branes of tension $\bar T_i(y_i) $. Let us
consider a particular case of the constant conformal
transformation \rf{b1} : $\Omega \equiv \const $. Then, with the
help of eqs. \rf{b2} and \rf{b3} for the conformally transformed
metric $g^{(D)}$ we obtain:
%%%%%%%%%%%%
\ba{b5} &{}&R_{MN}[g^{(D)}] -\frac12 g^{(D)}_{MN} R [ g^{(D)}] =
\\ &-& \bar \Lambda_D \Omega^2 g^{(D)}_{MN} -
\frac{\kappa^2_D}{\sqrt{| g^{(D)}|}} \sum _{i =0}^{n-1}\bar
T_i(y_i)\Omega^{-D+D_0+2} \sqrt{| g^{(D_0)}(x,y_i)|}\, \,
g^{(D_0)}_{\mu\nu}(x,y_i)
\delta_M^{\mu}\delta_N^{\nu}\delta(y-y_i)\, .\nn\ea
%%%%%%%%%%%%
This equation shows that conformally transformed metric $g^{(D)}$
is the solution for the model with the cosmological constant
$\Lambda_D \equiv \Omega^2 \bar \Lambda_D$ and the brane tensions
$ T_i(y_i) \equiv \Omega^{(-D+D_0+2)}\bar T_i(y_i)$. The latter
one is invariant for $D = D_0 +2$ which certainly is not the case
for $D=5$ if $D_0 =4$. Thus, if we want that solution
corresponding to a minimum of the effective potential for the
conformal excitations describes the model with the original
cosmological constant and the brane tensions, this minimum should
take place at $\Omega = 1$.

Let us consider transformation of the trace of the extrinsic
curvature evoked by conformal transformation of the metric. In the
case of 5-D metric \rf{2.5} written in Gaussian normal coordinates
the trace of the extrinsic curvature of the hypersurface $\Sigma:
r=r_i = \const$ reads
%%%%%%%%%%
\be{b6} K (r_i) = -\left. \nabla_{M}\, n^M\right|_{r_i} = - \left.
\frac{1}{2} g^{(4)\mu\nu} \frac{\partial
g^{(4)}_{\mu\nu}}{\partial r}\right|_{r_i} = - \left. 4
\frac{1}{a} \frac{d a}{d r}\right|_{r_i}\, ,\ee
%%%%%%%%%%%
where $n^M =\delta^M_r$ is the unit vector field orthogonal to
$\Sigma$. If the extrinsic curvature has a jump at this
hypersurface: $\widehat K (r_i) \equiv K (r_i^+) - K (r_i^-) \ne
0$ then it results in the Lanczos-Israel junction condition:
%%%%%%%%%%
\be{b7} T (r_i) = \frac{1}{\kappa^2_D} \frac34 \widehat K (r_i)\,
, \ee
%%%%%%%%%%
where $T(r_i)$ is the tension of the brane which causes the jump
of the extrinsic curvature.

For the metric $\bar g^{(D)}\, $, obtained with the help of the
conformal transformation \rf{b1} of the metric \rf{2.5}, the unit
vector field orthogonal to $\Sigma$ is $\bar n^M = \Omega^{-1}
\delta^M_r \Longrightarrow \bar n_M = \Omega \delta^r_M$. Here, we
consider the case when $\Omega = \Omega (x)$ does not depend on
the extra dimension $r$. Then, we obtain for the trace of the
extrinsic curvature of the conformal space-time:
%%%%%%%%%%%
\be{b8} \bar K (r_i) = -\left. \bar \nabla_{M}\, \bar
n^M\right|_{r_i} = - \left. \frac{4}{\Omega} \frac{1}{a} \frac{d
a}{d r}\right|_{r_i}\, .\ee
%%%%%%%%%%%
Correspondingly, the tensions of the brane in conformal and
original space-times are connected with each other as follows:
$\bar T (r_i) = \Omega^{-1} T (r_i)$ in accordance with equation
\rf{b5}.

%%%%%%%%%%%%%%%%%%%%%%%%%%%%%%%%%%%%%%%%%%%%%%%%%%%%%%%%%%%%%%%%%%%%
%\begin{appendix}
\bigskip

\section{Truncated conformal transformation  \label{appendix c}}
\setcounter{equation}{0}

\bigskip

In this appendix we shall show that our results do not change if
only additional dimension undergoes conformal perturbations in
metric \rf{2.5}:
%%%%%%%%%%%%%%
\be{c1} g^{(5)}(X) \Longrightarrow \bar g^{(5)}(X) = \Omega^2 (x)
dr\otimes dr + a^2(r)\gamma^{(4)}_{\mu \nu}(x)dx^{\mu}\otimes
dx^{\nu}\, .\ee
%%%%%%%%%%%%%
Subsequent application of appropriate formulas from Appendices A
and B yields
%%%%%%%%%%%%
\be{c2} \sqrt{|\bar g^{(5)}|}\, R[\bar g^{(5)}] = \Omega a^4\;
\sqrt{|\gamma^{(4)}|} \left\{ a^{-2} \left[ R[\gamma^{(4)}] - 2
\Omega^{-1}\Omega_{;\mu ;\nu} \gamma^{(4)\mu \nu} \right] -
\Omega^{-2} f_1(r) \right\} \, ,\ee
%%%%%%%%%%%
where $f_1(r)$ is defined in \rf{2.7}. To get this expression, it
is useful to go first to a new coordinate $R: dR = a^{-1}(r)dr$
and then, after using conformal transformation formulas, come back
to $r$ again. It can be easily seen that after conformal
transformation to the Einstein frame:
%%%%%%%%%%
\be{c3} \gamma^{(4)}_{\mu\nu} (x) \Longrightarrow \tilde
\gamma^{(4)}_{\mu\nu} (x) = \Omega (x)\gamma^{(4)}_{\mu\nu} (x)\,
\ee
%%%%%%%%%
and the dimensional reduction, action \rf{3.1} is exactly reduced
to effective action \rf{3.4}. Thus, gravexcitons have exactly the
same masses \rf{3.7} - \rf{3.11}. This result shows that geometry
(gravitational field) under conformal transformations behaves as
an elastic media. For an elastic body the eigen frequencies of its
oscillations do not depend on the manner of excitation.
%From other
%hand, it indicates that the results of our investigation are gauge
%invariant because we obtained the same physical result for two
%different gauges.

%%%%%%%%%%%%%%%%
\end{appendix}

%%%%%%%%%%%%%%%%%%%%%%%%%%%%%%%%%%%%%%%%%%%%%%%%%%%%%%%%

\end{document}